\documentclass{article}

\usepackage{arxiv}

\usepackage[utf8]{inputenc} 
\usepackage[T1]{fontenc}    
\usepackage{hyperref}       
\usepackage{url}            
\usepackage{booktabs}       
\usepackage{amsfonts}       
\usepackage{nicefrac}       
\usepackage{microtype}      
\usepackage{lipsum}
\usepackage{graphicx}
\graphicspath{ {./images/} }
\usepackage{subcaption}

\usepackage{hyperref}
\usepackage[capitalize]{cleveref}
\usepackage{siunitx}
\usepackage{endnotes}

\usepackage[backend=biber, maxbibnames=3]{biblatex}
\usepackage{xcolor}
\usepackage{tcolorbox}
\usepackage{fontawesome5}
\usepackage{listings}
\usepackage[margin=1in]{geometry}
\usepackage{tikz}
\usetikzlibrary{calc}

\definecolor{usercolor}{RGB}{240, 242, 245}
\definecolor{assistantcolor}{RGB}{240, 242, 245}
\definecolor{toolcallcolor}{RGB}{243, 240, 255}
\definecolor{toolresultcolor}{RGB}{240, 253, 244}
\definecolor{systemcolor}{RGB}{254, 242, 242}

\definecolor{useraccentcolor}{RGB}{71, 85, 105}
\definecolor{assistantaccentcolor}{RGB}{71, 85, 105}
\definecolor{toolaccentcolor}{RGB}{109, 40, 217}
\definecolor{resultaccentcolor}{RGB}{5, 150, 105}
\definecolor{systemaccentcolor}{RGB}{220, 38, 38}

\definecolor{ctxlow}{RGB}{220, 38, 38}
\definecolor{ctxmid}{RGB}{234, 179, 8}
\definecolor{ctxhigh}{RGB}{5, 150, 105}

\newlength{\chatwidth}
\newlength{\sidewidth}
\newlength{\colsep}
\newcommand{\chatrow}[2]{%
  \noindent
  \begin{minipage}[t]{\chatwidth}%
    \vspace{0pt}%
    #1%
  \end{minipage}%
  \hspace{\colsep}%
  \begin{minipage}[t]{\sidewidth}%
    \vspace{0pt}%
    #2%
  \end{minipage}%
  \par\vspace{2pt}%
}

\newcommand{\contextbar}[3]{%
  \pgfmathsetmacro{\fillratio}{#1/#2}%
  \pgfmathparse{\fillratio < 0.34 ? 1 : 0}\ifnum\pgfmathresult=1
    \colorlet{barfill}{ctxlow}
  \else
    \pgfmathparse{\fillratio < 0.67 ? 1 : 0}\ifnum\pgfmathresult=1
      \colorlet{barfill}{ctxmid}
    \else
      \colorlet{barfill}{ctxhigh}
    \fi
  \fi
  \pgfmathsetmacro{\pct}{int(\fillratio*100)}%
  \pgfmathsetmacro{\availw}{\sidewidth - 8}%
  \pgfmathsetmacro{\seggap}{2}%
  \pgfmathsetmacro{\segw}{(\availw - (#2-1)*\seggap) / #2}%
  \pgfmathsetmacro{\segh}{14}%
  \pgfmathsetmacro{\totalw}{#2*(\segw+\seggap)-\seggap}%
  \pgfmathsetmacro{\midx}{\totalw/2}%
  \resizebox{\linewidth}{!}{%
    \begin{tikzpicture}[baseline=(current bounding box.north), x=1pt, y=1pt]
      \foreach \i in {1,...,#2}{%
        \pgfmathsetmacro{\xpos}{(\i-1)*(\segw+\seggap)}%
        \pgfmathparse{\i <= #1 ? 1 : 0}%
        \ifnum\pgfmathresult=1
          \fill[barfill] (\xpos,0) rectangle (\xpos+\segw,\segh);%
        \else
          \fill[gray!20] (\xpos,0) rectangle (\xpos+\segw,\segh);%
        \fi
      }%
      \node[anchor=north, font=\small, text=gray]
        at (\midx, -1) {#3};%
    \end{tikzpicture}%
  }%
  \vspace{4pt}%
}

\newcommand{\sidecomment}[1]{%
  \begin{tcolorbox}[
    colback=gray!8,
    colframe=gray!25,
    boxrule=0.4pt,
    arc=4pt,
    left=4pt, right=4pt, top=3pt, bottom=3pt,
    width=\linewidth,
    fontupper=\small,
    nobeforeafter,
    box align=top
  ]
  \textcolor{gray}{#1}
  \end{tcolorbox}%
  \vspace{3pt}%
}

\newenvironment{conversationbox}{%
  \noindent\begin{tcolorbox}[
    colback=white,
    colframe=gray!30,
    boxrule=1pt,
    arc=8pt,
    left=8pt, right=8pt, top=8pt, bottom=8pt,
    nobeforeafter
  ]%
  \setlength{\sidewidth}{0.28\linewidth}%
  \setlength{\chatwidth}{\linewidth}%
  \addtolength{\chatwidth}{-\sidewidth}%
  \addtolength{\chatwidth}{-\colsep}%
  \noindent
  \begin{minipage}[b]{\chatwidth}%
    \textbf{Chat Messages}%
  \end{minipage}%
  \hspace{\colsep}%
  \begin{minipage}[b]{\sidewidth}%
    \textbf{Author’s Comments}%
  \end{minipage}%
  \par\vspace{4pt}%
  \noindent\rule{\linewidth}{0.4pt}%
  \vspace{6pt}%
}{%
  \end{tcolorbox}%
}

\newcommand{\usermsg}[1]{%
  \begin{tcolorbox}[
    colback=usercolor!70, colframe=useraccentcolor!30,
    boxrule=0.8pt, arc=6pt,
    left=6pt, right=6pt, top=4pt, bottom=4pt, boxsep=0pt,
    width=\linewidth, nobeforeafter
  ]
  \textcolor{useraccentcolor}{\textbf{\faUser\ User:}} #1
  \end{tcolorbox}%
}

\newcommand{\assistantmsg}[1]{%
  \begin{tcolorbox}[
    colback=white, colframe=assistantaccentcolor!40,
    boxrule=0.5pt, arc=6pt,
    left=6pt, right=6pt, top=4pt, bottom=4pt, boxsep=0pt,
    width=\linewidth, nobeforeafter
  ]
  \textcolor{assistantaccentcolor}{\textbf{\faComment\ Response:}} #1
  \end{tcolorbox}%
}

\newcommand{\toolcall}[2]{%
  \begin{tcolorbox}[
    colback=toolcallcolor, colframe=toolaccentcolor!30,
    boxrule=0.5pt, arc=5pt,
    left=6pt, right=6pt, top=4pt, bottom=4pt, boxsep=0pt,
    width=\linewidth, nobeforeafter
  ]
  \textcolor{toolaccentcolor}{\textbf{\faTools\ Tool:}} \texttt{#1}\\
  \textcolor{gray}{\textit{Parameters:}} #2
  \end{tcolorbox}%
}

\newcommand{\toolresult}[1]{%
  \hspace{10pt}\begin{minipage}{\dimexpr\linewidth-10pt}%
  \begin{tcolorbox}[
    colback=toolresultcolor, colframe=resultaccentcolor!30,
    boxrule=0.5pt, arc=5pt,
    left=6pt, right=6pt, top=4pt, bottom=4pt, boxsep=0pt,
    nobeforeafter
  ]
  \textcolor{resultaccentcolor}{\textbf{\faCheckCircle\ Result:}} #1
  \end{tcolorbox}%
  \end{minipage}%
}

\newcommand{\systemmsg}[1]{%
  \begin{tcolorbox}[
    colback=systemcolor, colframe=systemaccentcolor!30,
    boxrule=0.5pt, arc=6pt,
    left=6pt, right=6pt, top=4pt, bottom=4pt, boxsep=0pt,
    width=\linewidth, nobeforeafter
  ]
  \textcolor{systemaccentcolor}{\textbf{\faCog\ System:}} #1
  \end{tcolorbox}%
}

\newcommand{\contexticons}[2]{%
  \pgfmathsetmacro{\fillratio}{#1/#2}%
  \pgfmathparse{\fillratio < 0.34 ? 1 : 0}\ifnum\pgfmathresult=1
    \colorlet{iconfill}{ctxlow}
  \else
    \pgfmathparse{\fillratio < 0.67 ? 1 : 0}\ifnum\pgfmathresult=1
      \colorlet{iconfill}{ctxmid}
    \else
      \colorlet{iconfill}{ctxhigh}
    \fi
  \fi
  \noindent\small\textcolor{gray}{Context:\hspace{4pt}}%
  \foreach \i in {1,...,#2}{%
    \pgfmathparse{\i <= #1 ? 1 : 0}%
    \ifnum\pgfmathresult=1
      \textcolor{iconfill}{\large\faLightbulb}%
    \else
      \textcolor{gray!40}{\large\faLightbulb[regular]}%
    \fi
    \hspace{2pt}%
  }%
  \vspace{4pt}%
}

\title{KadiAssistant: A conversational AI Agent for information retrieval in Kadi4Mat}
\author{
Adrian Cierpka \\
  Karlsruhe Institute of Technology\\
  \texttt{adrian.cierpka@kit.edu} \\
\And
Mohammad Shafiqul Islam \\
Karlsruhe Institute of Technology\\
\texttt{mohammad.islam@kit.edu} \\
\And
Johannes Steinhülb\\
Karlsruhe Institute of Technology\\
\texttt{johannes.steinhuelb@kit.edu} \\
\AND
Eric Dietriche Sesso Domtchoueng\\
Karlsruhe Institute of Technology\\
\texttt{eric.domtchoueng@kit.edu} \\
\And
Michael Selzer\\
Karlsruhe Institute of Technology\\
\texttt{michael.selzer@kit.edu} \\
\And
Arnd Koeppe\\
Karlsruhe Institute of Technology\\
\texttt{arnd.koeppe@kit.edu} \\
}
\begin{document}
\maketitle
\begin{abstract}

We introduce \emph{KadiAssistant}, a privacy-by-design AI assistant integrated into the Kadi research data ecosystem, enabling researchers to efficiently access, aggregate, and synthesize information from heterogeneous, privacy-sensitive, and graph-structured research data. 
Interdisciplinary fields such as materials science bring together disciplines with their own terminology and standards. While this convergence fuels innovation, it also makes it increasingly difficult to connect and access knowledge, as data are distributed across disciplines, organizations, and individuals. Efficiently storing and sharing such heterogeneous data via research data platforms, such as Kadi4Mat, demands domain knowledge, technical expertise, and familiarity with metadata schemas and interfaces.
Research data also vary in sensitivity: newly generated `warm' data are often private, whereas published `cold' data are usually openly accessible. The Kadi ecosystem offers fine-grained access control for sensitive data. A solution for efficient information retrieval in Kadi must therefore respect the fine-grained access permissions.
To address these challenges, KadiAssistant combines a self-hosted large language model (LLM) with privacy-preserving semantic similarity search for Kadi4Mat into an agentic AI system. It autonomously plans and executes multiple retrieval steps, calling tools for semantic similarity search, metadata lookup, and record connections to screen, aggregate, and structure the information found into a highly informative answer. We evaluate KadiAssistant on a benchmark spanning three research datasets, showing better answer and source quality metrics than a single-step RAG-like baseline. Furthermore, we demonstrate that, for most configurations, the underlying k-NN search scales favorably to Kadi instances with up to one million vectors.
\end{abstract}

\section{Introduction}

Modern research generates vast amounts of heterogeneous data across multidisciplinary research data repositories. Even when FAIR data principles \cite{wilkinsonFAIRGuidingPrinciples2016a} are applied, effective information retrieval and data exploration may demand domain knowledge, technical expertise, and familiarity with diverse metadata schemas and interfaces.
This creates a high barrier to effectively using the repository and diverts valuable time from core research tasks. AI tools, especially large language model (LLM)-driven AI assistants, offer a promising path to lowering this barrier. 

Research data, however, vary in sensitivity over their lifecycle. We adopt the data lifecycle-temperature metaphor from Strange and Gooch \cite{strangeTakingTemperatureExploring2024}, dividing data into \emph{hot}, \emph{warm}, and \emph{cold} stages. \emph{Hot} data is newly generated, subject to change, and usually unpublished; \emph{warm} data remains in use but is updated infrequently; \emph{cold} data is in an archival state, likely published and rarely used or changed. Data repositories like Kadi4Mat that handle hot and warm data complicate the integration of AI assistants, as the AI assistant must respect fine-grained access control while operating on heterogeneous, frequently changing data. Repository platforms like Zenodo, by contrast, focus primarily on cold data, where access control is less of a concern.

Retrieval-augmented generation (RAG) \cite{lewisRetrievalAugmentedGenerationKnowledgeIntensive2021} has emerged as a common approach for enabling LLMs to answer questions using external data sources, like research data repositories. RAG is a three-step pipeline that first retrieves relevant data based on the question using a semantic similarity search, then augments the prompt with the found information, and finally generates an answer using an LLM. The semantic search relies on an embedding model that transforms the question into a high-dimensional vector, which it uses for a k-NN search in a vector database to find relevant embedded text chunks. 
More recently, a more flexible approach has emerged: agentic AI for information retrieval. In contrast to simple RAG pipelines, agentic systems allow the LLM to dynamically plan and execute multiple retrieval steps to gather information as needed \cite{xiSurveyLargeLanguage2026}. \Cref{sec:LLMs_RAG_AAI} explains the RAG pipeline and agentic AI in more detail. 

Several AI-enabled approaches have emerged to support analysis and retrieval of data from research data repositories. Sbail\`o et al. presented an AI toolbox based on Jupyter Notebooks that enables users to analyze data stored in \emph{NOMAD} \cite{sbailoNOMADArtificialIntelligenceToolkit2022} using artificial intelligence and machine learning. Kadi4Mat \cite{brandtKadi4MatResearchData2021} provides AI modules, such as KadiAI and CIDS \cite{cids}, to create and execute machine-learning processes \cite{rajagopalDataDrivenVirtualMaterial2023}. The goal of project AIRDEC is to improve the deposit and curation experience on Zenodo \cite{https://doi.org/10.25495/7gxk-rd71} using AI tools \cite{thenavigationfundAIRDECAlassistedRepository2025}.

Zhao et al. \cite{zhaoLISALithiumIonSolidState2025} have shown that their AI assistant LISA can answer expert questions about solid-state battery research. LISA uses a RAG architecture to retrieve data from a knowledge repository containing 160 documents, most of which are solid-state battery-related papers. Additionally, their work includes a feature called KadiChat that lets the user select data from Kadi4Mat to add to the assistant's knowledge base. However, KadiChat can only answer questions about three preselected documents, greatly limiting its ability to perform large-scale information retrieval from Kadi4Mat.
Clark et al. \cite{clarkSemanticResourcesManaging2025} published the Battery Knowledge Base, which uses LLMs and RAG to enable users to query for data.

AI-based information retrieval is also offered by commercial products such as ChatGPT \cite{openai_chatgpt_product} or Perplexity AI \cite{perplexity_ai_product}.
Both products can search the web, gather information, combine, condense, and structure it into an informative answer. However, these systems cannot access certain data repositories, as these may store unpublished or private data.

Recent work has recognized the need to incorporate access control policies into retrieval-augmented generation systems. Chen et al. \cite{chenIntegratingAccessControl2025} presented a proof-of-concept RAG system with fine-grained access control for sensitive patient data. Jeong and Lee \cite{jeongPermissionAwareRAGIdentity2025} proposed a novel permission-aware RAG framework that performs retrieval across heterogeneous multi-resource environments. More recently, Zhong et al. \cite{zhongHONEYBEEEfficientRolebased2026} introduced their framework \textit{Honeybee}, which uses RBAC-aware (Role-Based Access Control) dynamic partitioning to balance memory and retrieval efficiency in access-controlled vector search.

Access control addresses who is authorized to retrieve data, but integrating LLMs into research data repositories introduces an additional privacy concern: how the LLM and its provider handle retrieved data. This concern is especially relevant to data repositories handling hot and warm data, like Kadi.
As King et al. \cite{kingUserPrivacyLarge2025} reported, all of the six US-based LLM providers investigated train on user chat data by default, with limited opt-out options, and retain chat data for extended periods, sometimes indefinitely. Combined with known vulnerabilities that can facilitate the extraction of training data \cite{chenSurveyPrivacyRisks2025}, these practices heighten privacy concerns. We therefore advocate handling data within the organization using self-hosted LLMs and minimizing data exposure to external services.

We consider the integration of agentic AI with access-controlled semantic similarity search into a research data repository with existing access control policies to remain insufficiently explored. This leads to three research questions that guide our work:
\begin{description}
    \item[RQ1] Can we implement a semantic similarity search directly in Kadi's underlying database while respecting Kadi's fine-grained access-control rules?
    \item[RQ2] How do the query time of the k-NN search and the memory consumption of Kadi4Mat's vector table scale with the number of indexed vectors and records?
    \item[RQ3] Does a multi-step retrieval agentic architecture improve retrieval and answer quality over a single-step RAG-like baseline for a benchmark on Kadi? And how does this advantage change for varying retrieval complexity?
\end{description}

This work introduces \emph{KadiAssistant}, a privacy-by-design AI assistant for Kadi4Mat \cite{brandtKadi4MatResearchData2021} that tackles the intertwined problems of data retrieval, privacy concerns, and fine-grained access control. Our AI assistant enables natural-language-based information retrieval from documents and records metadata using LLMs, agentic AI, and a RAG-inspired semantic similarity search for Kadi4Mat. In contrast to existing work about permission-controlled RAG, we do not aim to introduce a new RAG framework or propose a new optimized vector retrieval scheme. Instead, KadiAssistant integrates permission-aware semantic retrieval into Kadi4Mat's underlying database by using the platform's existing fine-grained access-control logic. We view this new AI assistant as an important tool for improving information retrieval, lowering barriers to the efficient use of data repositories, and making a considerable contribution to the \emph{Findable} aspect of the FAIR data principles \cite{wilkinsonFAIRGuidingPrinciples2016a}.

\section{Technical Foundations}\label{sec:technical_foundations}
\subsection{The data repository: Kadi4Mat}
Modern research generates large, heterogeneous datasets from high-throughput experiments, simulations, and automated instrumentation. To manage and share this data, many communities rely on research data platforms such as NOMAD \cite{scheidgenNOMADDistributedWebbased2023}, The Materials Project \cite{10.1063/1.4812323, hortonAcceleratedDatadrivenMaterials2025}, and Kadi4Mat \cite{brandtKadi4MatResearchData2021}. These platforms aim to make data \emph{Findable}, \emph{Accessible}, \emph{Interoperable}, and \emph{Reusable} \cite{wilkinsonFAIRGuidingPrinciples2016a}.

Kadi4Mat \cite{brandtKadi4MatResearchData2021} is a general-purpose research data repository and electronic lab notebook supporting multiple research domains. Because the Kadi ecosystem focuses on hot and warm research data, it provides strong privacy and fine-grained access controls, ensuring that only authorized users or groups can access specific data.
Kadi4Mat organizes data around three core entities. \emph{Records} are the primary unit that couples descriptive metadata with attached content. \emph{Files}, such as PDFs, instrument logs, CSVs, and JSON files, store arbitrary data and are associated with a record. The third core entity, \emph{Collections}, groups together related records.

Record metadata is stored as key–value pairs and includes standard fields (e.g., identifier, title, description, creator) as well as an extensible “extras” field for domain-specific attributes.
Records can reference each other via annotated links, enabling users to represent relationships between records \cite{brandtKadi4MatResearchData2021}.

\subsection{RAG and agentic AI} \label{sec:LLMs_RAG_AAI}
AI assistants today typically rely on transformer-based \cite{vaswaniAttentionAllYou2023} large language models (LLMs), neural networks trained on large text datasets to generate human-like text. LLMs can summarize long documents, connect related information, and create answers using their broad general knowledge.

\begin{figure}
    \centering
    \includegraphics[width=0.9\linewidth]{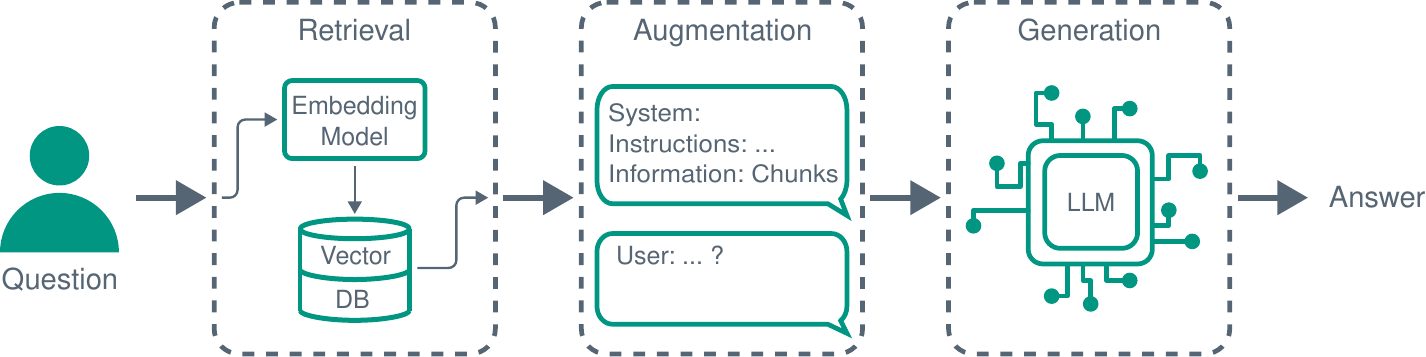}
    \caption{The RAG pipeline's three steps: retrieve information, augment the prompt, and generate an answer.}
    \label{fig:rag_concept}
\end{figure}

RAG \cite{lewisRetrievalAugmentedGenerationKnowledgeIntensive2021}, named after its three steps Retrieve, Augment, and Generate, emerged as a popular approach to give LLMs access to large data repositories. For a given user prompt, the three steps of RAG are executed:
\begin{enumerate}
    \item \emph{Retrieve}: An embedding model maps the query into a semantic vector space. Then the k nearest text chunks in this vector space are retrieved as relevant context. For efficient k-nearest neighbor (k-NN) search, the vectors are stored and queried in a vector database.
    
    \item \emph{Augment}: Construct a prompt that includes the user's query and the retrieved text chunks.
    \item \emph{Generate}: Feed the prompt into a large language model, which generates an answer to the user's question based on the data retrieved from the repository.
\end{enumerate}
\Cref{fig:rag_concept} visualizes this process.

To retrieve data from the vector database, it must be populated with data from the repository. The repository's data are split into chunks, mapped into the semantic vector space, and stored with their corresponding text in the vector database.

As powerful as this single-stage, k-NN-based semantic retrieval procedure is, adding a reranking stage after the initial data retrieval phase can improve retrieval quality \cite{zhaoLISALithiumIonSolidState2025, thakurBEIRHeterogenousBenchmark2021}.
In contrast to the first-stage bi-encoders, which create vector embeddings for query and text chunks independently, reranker models take both the query and text chunks into account to compute a similarity score. As the computational cost of scoring all text chunks in the repository scales linearly with their number, the reranker only scores the preselected chunks from the k-NN search.

In contrast to the static workflow of simple RAG pipelines, agentic systems allow the LLM to dynamically plan and execute multiple retrieval steps to gather information as needed \cite{xiSurveyLargeLanguage2026}.
Agentic systems can have a predefined workflow that, together with prompting, defines agent behavior. LangGraph \cite{langgraph}, for example, a framework for agentic AI, uses a state graph that the developer implements, defining fixed steps for the AI agent while also allowing decisions via conditional edges.
Another key extension of agentic AI is tools: functions the LLM can call. They are the interface that allows the AI agent to interact with the outside world. These tools can be used for information retrieval, e.g., a web search engine or a semantic similarity search for a research data platform.

\section{KadiAssistant: The components} \label{sec:concept}
\subsection{Overview and KadiWeb UI}
\begin{figure}[h]
    \centering
    \includegraphics[
        width=1.0\textwidth,
        alt={A depiction of the 6 components of KadiAssistant and how they are connected.}
        ]{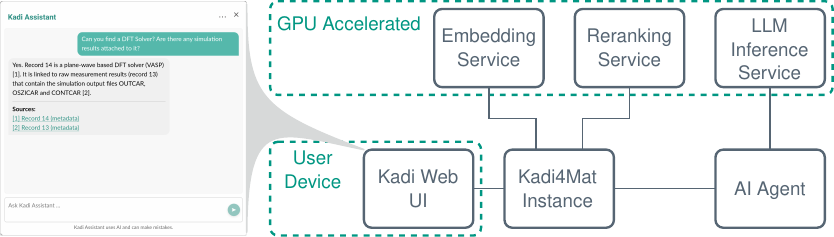}
    \caption{The overall architecture of the system. Kadi4Mat now has access to embedding and reranking services for the semantic similarity search. Other new components are the AI agent and the LLM inference service. The KadiWeb UI runs on the user's device, and the Embedding, Reranking, and LLM Inference Services are GPU-accelerated.}
    \label{fig:architecture}
\end{figure}

KadiAssistant consists of several components, depicted in \Cref{fig:architecture}. 
This division into components offers two direct advantages: it makes scaling the application easier, as components can be scaled individually, and it improves reliability, as the failure of one system (e.g., the reranking service) does not compromise the Kadi core functionality. 

We integrated a new chat box into KadiWeb, the web-based user interface for a Kadi instance. This allows the user to interact with KadiAssistant in a conversational style.
The chat box is depicted and described in \Cref{sec:demo}. When the user submits a message in the chat box, it is sent along with the chat history to the Kadi instance, which then forwards it to the AI agent component. 

\subsection{The AI agent and LLM inference server}
The AI agent runs on a dedicated component (\Cref{fig:architecture}) which defines the agent's workflow as a state graph, provides the tools, and specifies the prompts.
KadiAssistant has access to three tools: the semantic similarity search, a tool to retrieve metadata for a given record, and a tool to retrieve an object's (record or collection) connections to other objects.
The state graph, prompting, and tools are detailed in \Cref{sec:implAIAgent}.
The LLM runs on a GPU-accelerated LLM Inference Service. 

\subsection{Kadi4Mat with semantic similarity search} \label{sec:semsimseakadi}
The Kadi4Mat instance functions as our core database, connects the user interface to the AI agent, and now offers a new semantic similarity search that the AI agent can use. To add the semantic similarity search (as described in \Cref{sec:LLMs_RAG_AAI}) to Kadi4Mat, we extended our instance with an efficient k-nearest-neighbor (k-NN) search in the semantic vector space, an embedding service for vector creation, and a reranking service to score chunk similarities. 
Furthermore, Kadi4Mat now keeps the vector database synchronized with the records and files stored on the instance. 
The embedding and reranking models run on separate servers (\Cref{fig:architecture}), as they heavily benefit from GPU acceleration.
To make the new semantic search accessible, Kadi4Mat now exposes an interface that accepts a query and returns relevant text chunks.

For adding a k-NN search to a Kadi4Mat instance, we considered two approaches:
\begin{enumerate}
    \item A loosely coupled separate vector database on another server, or
    \item A tightly coupled vector database directly integrated into Kadi4Mat’s underlying database
\end{enumerate}
Running the vector database on a separate server allows us to choose from existing, performance-optimized, and easily scalable vector databases, such as ChromaDB \cite{chroma_db}, FAISS \cite{douze2025faisslibrary}, or Qdrant \cite{qdrant_db}. 
However, this loosely coupled design comes with a challenge, directly stemming from Kadi4Mat’s fine-grained access control: because the vector database stores vectors and text chunks from all users, it is necessary to protect them from unauthorized access by other users. This access rights logic already exists in Kadi4Mat, but cannot be directly accessed by a loosely coupled vector database server. Consequently, the access logic must either be duplicated or a sophisticated over-fetching algorithm with post-filtering must be developed. The duplication of access logic introduces redundancy and potential misalignment of access rights if the rules change. The over-fetching algorithm introduces more complexity to the implementation.

We therefore decided to implement the vector database directly into Kadi4Mat's underlying database. 
This tightly coupled approach allows us to leverage Kadi4Mat's access-control logic, which should make the implementation easier, more maintainable, and more secure than the loosely coupled vector database approach. 

\begin{figure}
    \centering
    \includegraphics[
        width=0.85\textwidth,
        alt={A flow diagram of the semantic similarity search, depicting how the data flows between the different components of KadiAssistant.}
        ]{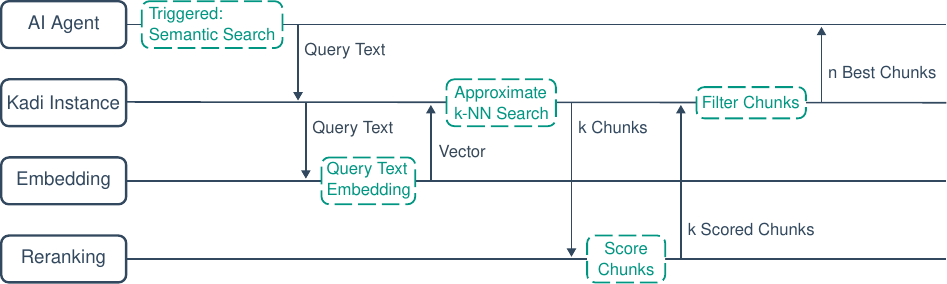}
    \caption{Flow diagram for the semantic similarity search. The Kadi instance leverages the embedding and reranking services to retrieve the n most similar chunks for the AI agent.} 
    \label{fig:rag_flow}
\end{figure}

\begin{figure}
    \centering
    \includegraphics[
        width=0.85\textwidth,
        alt={A flow diagram of the process that keeps the vector database synchronized with the data on the Kadi instance.}]{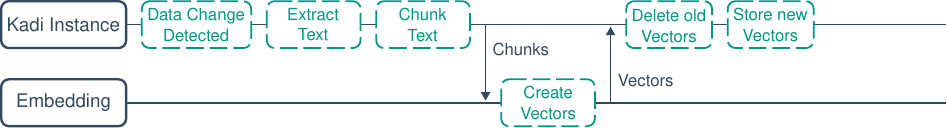}
    \caption{Flow diagram for synchronizing the vector database with data on the Kadi instance. For deletions, most steps are skipped, and the vectors are deleted.}
    \label{fig:file_change_flow}
\end{figure}

Two new important processes run on the Kadi4Mat instance: semantic similarity search and the vector database synchronization process that keeps its content up to date. \Cref{fig:rag_flow} depicts the components and processes involved in semantic similarity search. The update process of the vector database (\Cref{fig:file_change_flow}) is triggered when a data change is detected. For file or record deletions, most steps are skipped, and only the old vectors are deleted.

\section{Implementation Details}\label{sec:impl}

\subsection{Implementation of the AI agent and the user interface} \label{sec:implAIAgent}
The user interface of KadiAssistant is implemented as a plugin for Kadi4Mat. The chat history is saved locally in the user's browser rather than in the Kadi instance, enhancing user data privacy. When a user message is submitted, it is appended to the chat history. This updated history is sent via an HTTP request to the Kadi instance, which forwards it to the AI agent.

The AI agent’s logic is implemented using LangGraph \cite{langgraph} and packaged as a FastAPI server.
This server exposes an HTTP endpoint that accepts the chat history, the URL to the Kadi instance, and a user-specific OAuth access token and returns the AI assistant's answer. The URL and access token are needed so the AI assistant's tools can access the Kadi instance.

\begin{figure}
    \centering
    \includegraphics[
        width=0.9\textwidth,
        alt={A visualization of the state graph of the AI agent.}]{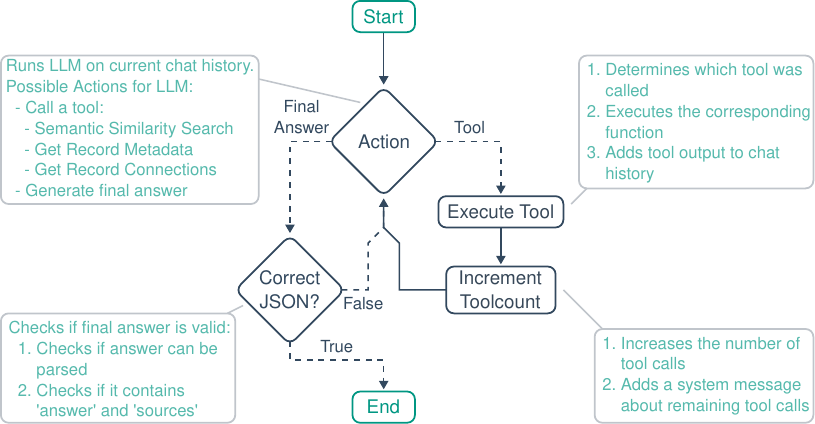}
    \caption{The state graph of our AI agent. The entry point is at \textit{Start}.} 
    \label{fig:kadichat2-graph}
\end{figure}

A new request to the HTTP endpoint triggers the answer-generation process. First, a system message that steers the AI agent's behavior is generated and pushed to the top of the chat history.
The program enters the AI agent's state graph, which defines the steps the agent can take to reach an answer (\Cref{fig:kadichat2-graph}). The graph allows the LLM either to call tools or to generate the final answer. After each tool execution, the LLM is informed about the remaining tool calls. When the LLM generates the final answer, it is checked against the desired template. If the answer matches the template, it is sent to the user; otherwise, the LLM generates a new answer.

KadiAssistant's behavior is heavily controlled by prompting. Most prompting is in the initial system message added at the beginning of the chat history. 
This message contains information about KadiAssistant itself, a description of Kadi4Mat, the rules KadiAssistant has to follow, and additional information.

Further prompting is located in smaller system messages and tool descriptions. The smaller system messages contain information about remaining tool calls or about violations of the answer format. 
To give KadiAssistant the ability to gather data from Kadi4Mat, we implemented three tools using KadiAPY \cite{kadiapy_kadi4mat_team_and_contributors_2025_15422101}, a Python library to interact with the Kadi instance: 
\begin{itemize}
    \item Semantic similarity search takes a query text as input and returns the eight most similar chunks.
    \item The ``get metadata'' tool retrieves the metadata for a given record ID.
    \item ``Get connections'' tool takes the ID of a Kadi4Mat object and the object type (``record'' or ``collection'') as input and returns the object's connections.
\end{itemize}
All three tools use a user-specific OAuth access token to only retrieve data from Kadi that the user may access. This is a key design decision to make sure KadiAssistant's data retrieval is private-by-construction.

\subsection{LLM, embedding, and reranker server}
As data privacy is a key feature of KadiAssistant, we host the LLM, embedding, and reranking components on our servers using vLLM \cite{kwonEfficientMemoryManagement2023_vLLM}, an open-source inference engine for large language models.
For the LLM, we use the model gpt-oss-120B \cite{openai2025gptoss120bgptoss20bmodel}.
As for the embedding and reranking models, we use \textit{Qwen3-Embedding-0.6B} and \textit{Qwen3-Reranker-0.6B} \cite{zhangQwen3EmbeddingAdvancing2025}, respectively, as these models offer a good trade-off between size and quality. Another benefit is that these models are multilingual, so if the user asks a question in English, they can retrieve and score chunks in other languages. Each model runs on an NVIDIA A100 80GB GPU.

\subsection{Kadi4Mat with k-NN-Vector search}
As a core component of KadiAssistant, we extended Kadi4Mat with new functionality.
We added new HTTP endpoints, including one for accessing KadiAssistant and one for the semantic similarity search, both only accessible to logged-in users. Kadi4Mat now also runs the process for updating the vector table, keeping it synchronized with Kadi’s data.

To add semantic similarity search to Kadi4Mat, we added the pgvector \cite{pgvector} extension and a new vector table to Kadi4Mat’s underlying PostgreSQL database, containing the vector embedding, the chunk text, and information that determines the text's source.
For an efficient k-NN search on the embedding column, we rely on the HNSW (Hierarchical Navigable Small World) index \cite{malkovEfficientRobustApproximate2018}, a data structure for fast approximate k-NN search.
The access rights filter for the semantic similarity search uses the record\_id column of the vector table and existing access control functions for records in Kadi4Mat.
Using the fields ``record\_id'', ``file\_name'', and ``from\_metadata'' in a vector table row, the AI assistant can identify a chunk's source. This matters because we want the AI assistant to provide the user with the sources its answer is based on.

The Kadi instance also handles synchronizing the vector table with the data stored on the instance. The general flow is shown in \Cref{fig:rag_flow}. This process is implemented as a background task using the Celery task queue, as we do not want this process to block the user's interaction with Kadi4Mat.
To extract the content from files and records and chunk it afterward, we used a variety of libraries, such as \textit{unstructured} \cite{unstructured} or LangChain \cite{langchain}, and implemented missing functionality ourselves. For example, we implemented a JSON splitter with truncation.

To ensure that the text field for vectors derived from record metadata does not accidentally leak data, we used only record metadata accessible to all users with read permission for the record. As an example, we did not store record connections in the chunks, as they may link to records the user might not have access to.

\section{Evaluation Setup} \label{sec:methods}
\subsection{Datasets} \label{sec:methods:datasets}
Our demonstration of KadiAssistant (\Cref{sec:demo}), parts of the scaling analysis (\Cref{sec:results:scaling}), and the quantitative analysis (\Cref{sec:results:quantitative}) are based on three example datasets:
\begin{itemize}
    \item LISA-Replica: This dataset is a replication of the dataset used in the LISA paper \cite{zhaoLISALithiumIonSolidState2025}. Our replication includes all 158 papers from the original dataset but omits the two websites. This is a natural-language-rich but metadata-poor dataset consisting of a single record.
    \item ML-Breathing-Detection: A metadata-heavy dataset created by Grolig et al. \cite{grolig_2023_8366576}. This ontology dataset comprises training and evaluation data for machine learning models for breathing detection using CIDS \cite{cids}, containing 417 highly connected records.
    \item POLiS Ontology: A dataset created by many contributors, based on the ontology created by Noman and Selzer \cite{nomanEnhancingMultiscaleSimulation2025}. This dataset is metadata-heavy and models the connections between records. In contrast to the ML-Breathing-Detection dataset, this dataset contains natural language, often in the record descriptions.
\end{itemize}
The POLiS Ontology datasets can be found on Zenodo \cite{cierpka_2026_22814081}. We did not publish the LISA-Replica and ML-Breathing-Detection datasets due to copyright and data sensitivity concerns.

\subsection{Test of the semantic similarity search access rights filter}
KadiAssistant is designed to be private-by-construction, including our newly implemented semantic similarity search. To investigate if the semantic similarity search obeys Kadi's access control rules (RQ1), we implemented several tests. These tests examine whether the semantic similarity search and the underlying new access filter return only vectors the user may access.
The tests include cases for private records, deleted records, and records with granted and revoked access permissions.

\subsection{Scalability experiments} \label{sec:methods:scalability}
As production instances of Kadi4Mat can be orders of magnitude larger than the three example datasets that we showcase in this paper, we need to predict how KadiAssistant scales to larger instances. The component most affected by the amount of data is the vector database.
We investigated two properties:
(1) the response time for the k-NN search up to one million vectors and (2) the memory consumption of the vector database. 

\subsubsection{Retrieval time scalability} \label{sec:exp_setup:retrieval_setup}
The newly implemented semantic similarity search consists of three main steps: 1. The query embedding, 2. The k-NN search, and 3. The reranking of chunks (see \Cref{fig:rag_flow}). As Steps 1 and 3 are not influenced by the size of the vector database and may introduce noise to the runtime measurements (through network latency and model response time), we decided to investigate the k-NN search only, rather than the whole semantic similarity search pipeline.

To investigate the runtime of the k-NN search, we conducted two experiments: one in which we kept the number of records constant and another in which it grew linearly with the number of vectors. We did so to investigate both the scaling of the HNSW index and the realistic growth of the database.
For both experiments, we measured the query time up to one million vectors sampled uniformly from a 1,024-dimensional (hyper-)sphere. We also modeled three access fractions (0.01, 0.1, and 1.0), corresponding to the fraction of vectors the k-NN search can access. 
Both experiments were executed on a virtual machine.

In the first experiment, we kept the number of records constant at two, one for all accessible vectors and one for all the non-accessible vectors. For each data point, we averaged the retrieval time over 100 trials (k-NN queries with k=50), and fitted a log function to each access fraction.

For the second experiment, we increased the number of records approximately linearly with the number of vectors by defining the maximum number of vectors per record. We chose three configurations: 10, 100, and 1000 vectors per record. For this experiment, each data point was created by averaging over 500 k-NN runs, again with k=50.

\subsubsection{Memory scalability} \label{sec:methods:scalability:memory}
As vector databases grow with the amount of data stored on a Kadi4Mat instance, we need to approximate memory consumption for larger databases.
To do so, we used four databases of varying sizes: the three we already introduced, plus one containing all the data from the other three. 
For each database, we measured the memory size of the HNSW index and of the complete vector table, including the table, HNSW index, and TOAST (The Oversized-Attribute Storage Technique) data. 

For this experiment, we expect a linear scaling of both the HNSW and the total vector table memory.
\textcite{malkovEfficientRobustApproximate2018} showed theoretically and empirically that HNSW memory consumption grows as $\mathcal{O}(N)$, where $N$ is the number of vectors.
As the TOAST data and the vector table itself should grow linearly with the number of vectors, we expect $\mathcal{O}(N)$ scaling for the total vector data. For both investigated properties, we fitted a linear function of the form $ax+b$ ($b \geq 0$) to predict further memory scaling.

\subsection{Quantitative answer experiments}

\subsubsection{Baseline models} \label{sec:methods:baseline_models}
We constructed three baseline models (\Cref{tab:baseline_models}). All models are variants of KadiAssistant and use the same underlying code and agent framework. This setup allows us to compare the effect of the number of tool calls and available tools without introducing differences caused by independently implemented baseline models.

KadiAssistant is configured with five tool calls and has access to all three tools. For the baseline models, we varied the available tools and the number of allowed tool calls.  The baseline model \textit{Agent-1TC-OSS} is limited to one tool call; we also restricted the tools to semantic similarity search. This configuration is intended to mimic a simple RAG pipeline with reranking, in which the LLM generates the query used for semantic search. For model \textit{Agent-1TC-AT}, we limited the number of tool calls to one but gave it access to all tools, as we are interested in how a model with only one tool call compares against KadiAssistant. For these two single-tool-call variants, we extended the ruleset of the system
prompt to instruct the LLM to use only one tool call (see supplementary material Section 1). We added the third variant \textit{Agent-3TC-AT} to the baseline models with three tool calls and access to all tools, to see how a model capable of multi-step retrieval with fewer tool calls compares against KadiAssistant.

\begin{table}[htbp]
\centering
\caption{KadiAssistant and baseline variants}
\label{tab:baseline_models}
\begin{tabular}{l ccc}
\toprule
Model & Tool calls & Tools & Purpose \\
\midrule
Agent-1TC-OSS & 1 & Semantic similarity search only & Single-step RAG-like baseline \\
Agent-1TC-AT & 1 & All tools & Isolate the effect of multi-step retrieval\\
Agent-3TC-AT & 3 & All tools & Examine effect of tool-call budget\\
KadiAssistant & 5 & All tools & Original system\\
\bottomrule
\end{tabular}
\end{table}

\subsubsection{Benchmark construction} \label{sec:methods:benchmark_construction}

To evaluate KadiAssistant and the baseline models quantitatively, we created a new benchmark based on the three datasets (see \Cref{sec:methods:datasets}). For each dataset, we created questions of varying complexity. 
For the ontology-based datasets, we control the question complexity using a metric we call hops. This metric describes the number of record links the AI assistant is intended to traverse to find the answer to the question. 
For the LISA dataset, the question complexity is controlled by a toggle to either include a hint to the paper the question stems from or not include a hint.
To construct this benchmark, we used a four-phase procedure: (1) Kadi database setup, (2) automated candidate generation for each dataset, (3) manual duplicate resolution, and (4) merging the three partial benchmarks into one dataset.
\paragraph{Phase 1: Kadi database setup.}
We created a new Kadi instance and uploaded the three datasets presented in \Cref{sec:methods:datasets}. This instance is used for benchmark construction and the evaluation of KadiAssistant and baseline models.
\paragraph{Phase 2: Automated generation.}
For each dataset, a generation script produces data points with different levels of complexity. Each of these data points consists of a question (gt-question) and a corresponding answer (gt-answer). It also includes the Kadi resources used to generate the question and answer (gt-full-sources), the Kadi resources cited in the answer (gt-answer-sources), and some additional information about the generation process. 
The three generation scripts differ from each other, but they all repeatedly execute the following four steps to generate a new data point.
\begin{enumerate}
    \item \emph{Data sampling}: The script samples a small set of Kadi resources (records or files).
    \item \emph{LLM-based generation}: Based on the sampled data, an LLM generates a question, answer, and, depending on the dataset, additional information.
    \item \emph{Duplicate flagging}: An LLM is prompted to find all existing data points that might duplicate the new one and store them with the new data point.
    \item \emph{Storage}: The new data point is stored in a partial benchmark dataset.
\end{enumerate}
For more details on the steps, see the supplementary material Section 2. 

\paragraph{Phase 3: Manual duplicate resolution.} 
We manually reviewed all potentially duplicated data points and decided which to keep and which to discard. To do this efficiently, we used a helper script that identifies the clusters of flagged duplicates and lets us decide what to discard and what to keep.

\paragraph{Phase 4: Merging.}
The three de-duplicated, per-dataset benchmarks were merged into one benchmark dataset. For later analysis of the evaluation results, we kept information about data point construction and the corresponding dataset.

We deliberately separated the automated flagging of duplicates and the manual decision process. We viewed duplicate detection as a cheap process that can be automated using an LLM, whereas deciding which duplicate to keep or discard is more complex and therefore requires manual oversight.

To reduce the potential influence of author bias, we implemented a highly automated generation process. Additionally, although LLMs are black boxes, the available source code and LLM prompts provide more explicit documentation of the generation process and facilitate the creation of similar datasets.
To generate the dataset, we used the model Qwen3.5-397B-A17B \cite{qwen3.5} hosted on servers of the SCC at KIT. However, this LLM-generated benchmark may not fully reflect the diversity or ambiguity of real researcher queries. We therefore consider the benchmark primarily as a basis for comparing KadiAssistant with the baseline models, rather than as a comprehensive representation of real-world queries.

\subsubsection{Benchmark statistics and generation details}
Our generated benchmark consists of 159 data points. 60 data points are from the LISA-Replica dataset, 53 from the ML-Breathing-Detection dataset, and 46 from the POLiS Ontology dataset. \Cref{tab:benchmark-stats} shows these benchmark statistics, including the number of questions per complexity setting.

\begin{table}[htbp]
\centering
\caption{Benchmark statistics: Requested number of data points and actual number in the dataset}
\label{tab:benchmark-stats}
\begin{tabular}{l ccc}
\toprule
Dataset & Complexity Name & Requested & Actual \\
\midrule
LISA-Replica & With Paper Hint & 30 & 30 \\
             & Without Paper Hint & 30 & 30 \\
\addlinespace
ML-Breathing-Detection & 0-Hop & 30 & 26\\
                       & 1-Hop & 30 & 27\\
\addlinespace
POLiS Ontology & 0-Hop & 20 & 19\\
               & 1-Hop & 20 & 13\\
               & 2-Hop & 20 & 14\\
\bottomrule
\end{tabular}
\end{table}

\subsubsection{Retrieval and answer-quality metrics}
To analyze the generated answer and sources quantitatively, we use six different metrics. All metrics are bounded between zero and one, and higher values indicate better performance.
The first three metrics are computed on the sources only (gt-full-sources, gt-answer-sources, and the AI assistant's sources: assistant-sources):
\emph{Recall} is calculated on the gt-answer-sources and assistant-sources. It measures the fraction of gt-answer-sources that the assistant cited. A lower recall indicates that the assistant failed to retrieve or cite some sources. $$\text{recall} = \frac{|\text{gt-answer-sources}\cap\text{assistant-sources}|}{|\text{gt-answer-sources}|}$$
\emph{Strict precision} is calculated on the gt-answer-sources, and the assistant-sources. It measures the fraction of assistant-sources that are actually part of the gt-answer-sources. A lower precision indicates that the assistant retrieved sources that are actually irrelevant or incorrect. $$\text{strict precision} = \frac{|\text{gt-answer-sources}\cap\text{assistant-sources}|}{|\text{assistant-sources}|}$$
\emph{Fuzzy precision} is calculated analogously to strict precision, but using gt-full-sources and assistant-sources. We added this metric because we do not want to penalize potentially relevant sources that are not cited in the gt-answer. This metric is more permissive than strict precision. $$\text{fuzzy precision} = \frac{|\text{gt-full-sources}\cap\text{assistant-sources}|}{|\text{assistant-sources}|}$$

To evaluate the answer texts, we rely on existing metrics implemented in RAGAS \cite{esRagasAutomatedEvaluation2025}, an evaluation framework to assess the performance of RAG systems. For our work, we use two metrics directly from RAGAS and adapt another one:
\emph{Answer relevancy} assesses how well the generated answer addresses the posed question. This metric penalizes incomplete answers and those with unnecessary details.
The \emph{answer correctness} metric from RAGAS measures similarity between the ground-truth answer and the generated answer.
Our \emph{source faithfulness} metric is based on the faithfulness metric from RAGAS. In RAGAS, it gauges how faithful the generated answer is to the context the AI assistant retrieved. In our case, however, we use this metric to measure how well the assistant's answer is supported by the sources the assistant cites. We do this by using the content of assistant-sources instead of the context that the AI assistant retrieved with its tools.

As some RAGAS metrics rely on an LLM or embedding model, we used gpt-oss-120B \cite{openai2025gptoss120bgptoss20bmodel} and Qwen3-Embedding-8B \cite{zhangQwen3EmbeddingAdvancing2025}, both hosted at SCC (KIT). The resulting RAGAS-based scores provide useful indications of answer quality and allow for quantitative comparison between systems. However, they should be interpreted with caution, as the capabilities and biases of the underlying models may influence them.

\section{Demonstration, Results and Discussion} \label{sec:results}
\subsection{Demonstration of KadiAssistant} \label{sec:demo}

\begin{figure}
    \centering
    \includegraphics[
        width=0.5\textwidth,
        alt={A depiction of the chatbox of KadiAssistant. The chat history shows a user question and KadiAssistant's answer. The answer consists of the answer text and the referenced sources below.}
        ]{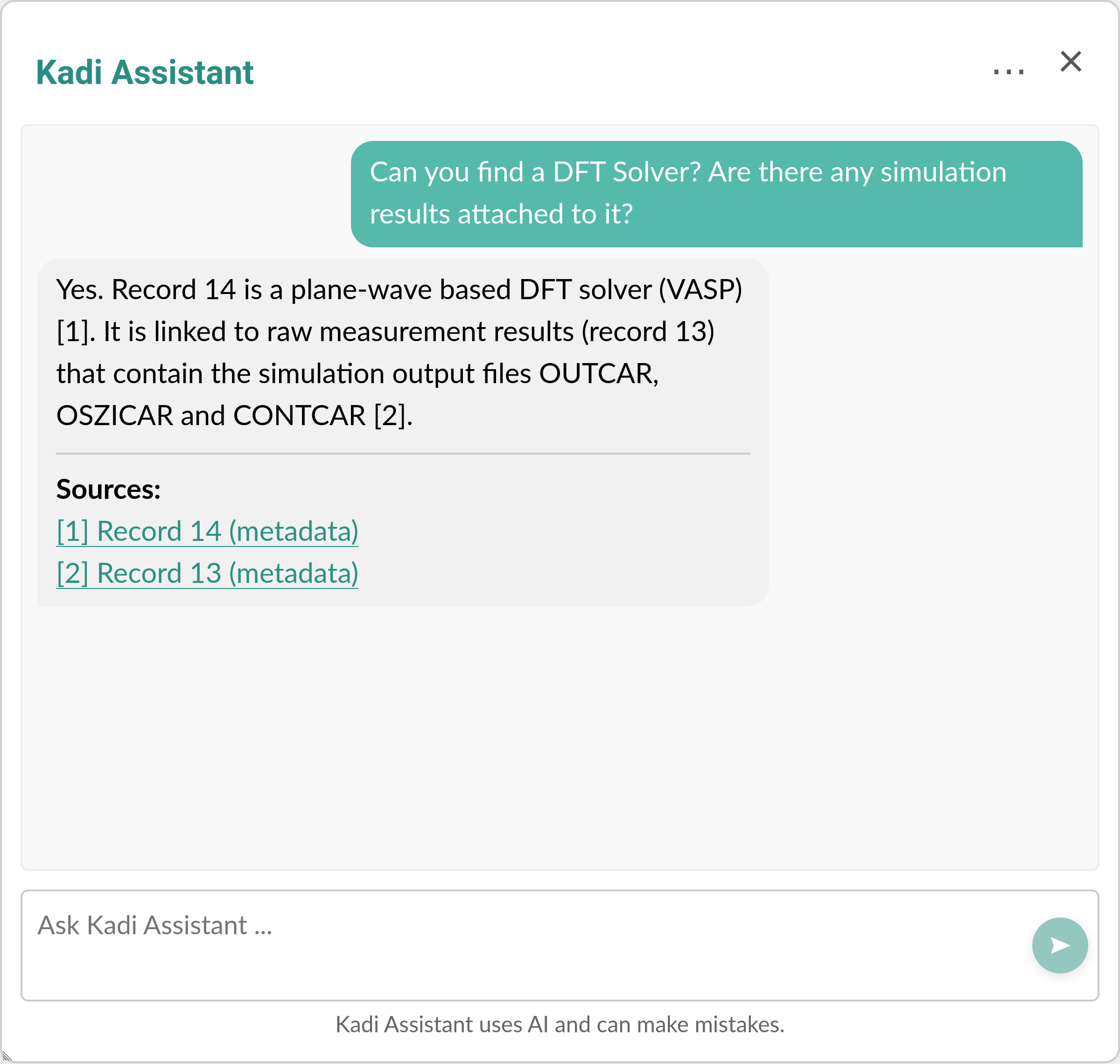}
    \caption{KadiAssistant's user interface. This user interface is integrated into KadiWeb. The AI assistant's response includes the answer, along with the sources used to generate it, directly below.}
    \label{fig:kadichat_ui}
\end{figure}

In this section, we demonstrate KadiAssistant's user interface and capabilities qualitatively on our three datasets (\Cref{sec:methods:datasets}).
Figure \ref{fig:kadichat_ui} shows the chat-based user interface of KadiAssistant, integrated into Kadi4Mat's web interface.
KadiAssistant's response comprises the answer and the sources used, providing the user with an entry point for validating the answer and further research. Each source includes the record and additional information identifying the specific source content. This can either be a file name if the source is a file, or ``metadata'' if the source is the record’s metadata.

\subsubsection{Example for the LISA-Replica dataset}
As a first example, we ask KadiAssistant to locate the paper titled ``Kadi4Mat: A research data infrastructure for materials science'' of the LISA-Replica dataset. This task should be solvable using only one tool call. This question is challenging as the file name ``1282-1-9024-1-10-20210210.pdf'' does not name or hint at Kadi4Mat at all, making it difficult for a user to locate. \Cref{fig:conversation-lisa-kadi} depicts the internal steps of KadiAssistant, including our comments on the most important steps, to help the reader interpret what is happening.

To locate the paper, the AI assistant calls the semantic similarity search and finds eight text chunks from different files. At least one chunk is from the correct file. However, some retrieved chunks are misleading. For example, one chunk comes from a paper that only cites the correct paper. The AI assistant correctly recommends the file ``1282-1-9024-1-10-20210210.pdf'' to the user. 

\begin{figure}[htbp]
  \centering
  \begin{conversationbox}
    \chatrow{\systemmsg{Your name is \ldots}}
            {\sidecomment{System prompt with important rules and base information.}}

    \chatrow{\usermsg{Can you locate the paper 'Kadi4Mat: A research data infrastructure for materials science' for me?}}
            {\sidecomment{User input is appended and the assistant starts information retrieval}}

    \chatrow{\toolcall{Kadi\_Similarity\_Search}{``text'': ``Kadi4Mat: A research data infrastructure for materials science''}}
            {}

    \chatrow{%
      \toolresult{$[\cdots]$ $k$ best ranked chunks, not shown}%
    }{%
      \sidecomment{Assistant found several files mentioning 'Kadi4Mat: A research data infrastructure for materials science'} \\
      \contexticons{1}{1}
    }

    \chatrow{\systemmsg{You have 4 tool call(s) remaining out of 5.}}{}
    
    \chatrow{%
      \assistantmsg{The paper is available in record 1 as the file ``1282-1-9024-1-10-20210210.pdf''.\\[4pt]
      \textit{Sources:}\\
      1. Record 1 (1282-1-9024-1-10-20210210.pdf)
      }%
    }{%
      \sidecomment{This answer is correct. KadiAssistant decided to cite the correct paper rather than another file that merely lists 'Kadi4Mat: A research data infrastructure for materials science' as a source.}%
    }
  \end{conversationbox}
  \caption{A deeper insight into an example for the LISA-Replica dataset. The AI assistant is tasked with finding the Kadi4Mat paper \cite{brandtKadi4MatResearchData2021} on Kadi. The difficulty of this task (especially for users) stems from the filename, which includes no hint of Kadi4Mat.}
  \label{fig:conversation-lisa-kadi}
\end{figure}

\subsubsection{Example for the POLiS Ontology dataset}
To demonstrate the capabilities of KadiAssistant's agentic architecture, we chose the question ``Can you find a DFT Solver? Are there any simulation results attached to it?'' on the POLiS Ontology dataset. This question should require multiple tool calls.
\Cref{fig:conversation-dft} depicts KadiAssistant's internal process. First, KadiAssistant uses semantic similarity search and finds record 14, which represents a DFT solver. In a second step, the assistant searches the record's connections and finds the relevant record 13, ``RawMeasurementResult''. With the last tool call, KadiAssistant screens the record's metadata and afterward creates the final answer.

\begin{figure}[htbp]
  \centering
  \begin{conversationbox}
    \chatrow{\systemmsg{Your name is \ldots}}
            {\sidecomment{System prompt with important rules and base information.}}

    \chatrow{\usermsg{Can you find a DFT Solver? Are there any simulation results attached to it?}}
            {\sidecomment{User input is appended and the assistant starts information retrieval}}

    \chatrow{\toolcall{Kadi\_Similarity\_Search}{``text'': ``DFT Solver''}}
            {}

    \chatrow{%
      \toolresult{$[\cdots]$ $k$ best ranked chunks, not shown}%
    }{%
      \sidecomment{Assistant found record 14, representing a DFT solver} \\
      \contexticons{1}{3}
    }

    \chatrow{\systemmsg{You have 4 tool call(s) remaining out of 5.}}{}

    \chatrow{\toolcall{Kadi\_Get\_Connections\_Tool}{``id'': 14, ``type'': ``record''}}
            {}

    \chatrow{%
      \toolresult{$\{\cdots\}$ Connections of record 14, not shown}%
    }{%
      \sidecomment{Assistant found a link mentioning Record 13: ``RawMeasurementResult''}%
      \contexticons{2}{3}
    }

    \chatrow{\systemmsg{You have 3 tool call(s) remaining out of 5.}}
    {}

    \chatrow{\toolcall{Kadi\_Get\_Meta\_Data\_Tool}{``record\_id'': 13}}
            {}

    \chatrow{%
      \toolresult{$\{\cdots\}$ Metadata of record 13, not shown}%
    }{%
      \sidecomment{Metadata of record 13 confirms that it contains simulation results}%
      \contexticons{3}{3}
    }

    \chatrow{\systemmsg{You have 2 tool call(s) remaining out of 5.}}
    {}
    
    \chatrow{%
      \assistantmsg{Yes. Record 14 is a plane‑wave based DFT solver (VASP) [1]. It is linked to raw measurement results (record 13) that contain the simulation output files OUTCAR, OSZICAR and CONTCAR [2].\\[4pt]
      \textit{Sources:}\\
      1. Record 14 (metadata)\\
      2. Record 13 (metadata)}%
    }{%
      \sidecomment{Answer generated from 2 sources.}%
    }
  \end{conversationbox}
  \caption{An example use case for the POLiS Ontology dataset necessitating multiple tool calls.}
  \label{fig:conversation-dft}
\end{figure}

\subsubsection{Example for the ML-Breathing-Detection dataset}
\label{sec:ml-breathing-example}
KadiAssistant can also navigate through Kadi4Mat resources when the semantic similarity search returns chunks of minor relevance. 
Initially, we built the following example to showcase a simple one-step task: KadiAssistant retrieving the dataset for the machine learning models in the ML-Breathing-Detection dataset. However, we found that the retrieved chunks from the semantic similarity search do not contain the dataset record itself; rather, they are less relevant chunks from records containing machine learning models. Despite this, KadiAssistant recovered by analyzing the links of one of the retrieved records. There, it found a record that represents the training procedure, which is itself linked to the dataset record. 
\Cref{fig:conversation-breathing-dataset} depicts an overview of the steps KadiAssistant took. 
For this dataset, the decisions of KadiAssistant led to the correct conclusion. In \Cref{sec:limitations_and_challenges}, we discuss a possible failure case.
\begin{figure}[htbp]
  \centering
  \begin{conversationbox}

    \chatrow{\usermsg{Please locate the dataset that the breathing detection models were trained on.}}
            {}

    \chatrow{\toolcall{Kadi\_Similarity\_Search}{``text'': ``breathing detection models dataset''}}
            {}

    \chatrow{%
      \toolresult{$[\cdots]$ $k$ best ranked chunks, not shown}%
    }{%
      \sidecomment{Assistant found eight chunks from breathing detection models, which, despite being scored as highly relevant, we view as sources of minor significance. One of them is record 369.}\\
      \contexticons{1}{4}
    }


    \chatrow{\toolcall{Kadi\_Get\_Connections\_Tool}{``id'': 369, ``type'': ``record''}}
            {}

    \chatrow{%
      \toolresult{$\{\cdots\}$ Connections of record 369, not shown}%
    }{%
      \sidecomment{Assistant found a link mentioning Record 279: ``cids-breathing-detection-results-training040''}%
      \contexticons{2}{4}
    }

    {}
    \chatrow{\toolcall{Kadi\_Get\_Connections\_Tool}{``id'': 279, ``type'': ``record''}}
            {}

    \chatrow{%
      \toolresult{$\{\cdots\}$ Connections of record 279, not shown}%
    }{%
      \sidecomment{Assistant found a link mentioning Record 3: ``cids-breathing-detection-tfrecords''}%
      \contexticons{3}{4}
    }

    {}
    
    \chatrow{\toolcall{Kadi\_Get\_Meta\_Data\_Tool}{``record\_id'': 3}}
            {}

    \chatrow{%
      \toolresult{$\{\cdots\}$ Metadata of record 3, not shown}%
    }{%
      \sidecomment{Metadata of record 3 is heavily truncated, as it contains many files. But it contains the key-value pair: 'type': ``kadiai:dataset''}%
      \contexticons{4}{4}
    }

    {}
    
    \chatrow{%
      \assistantmsg{The breathing detection models were trained on the dataset record with ID 3, identifier 'cids-breathing-detection-tfrecords\_'.\\[4pt]
      \textit{Sources:}\\
      1. Record 279 (metadata)\\
      2. Record 3 (metadata)}%
    }{%
      \sidecomment{The answer is correctly derived.}%
    }
  \end{conversationbox}
  \caption{An example task for KadiAssistant on the ML-Breathing-Detection dataset. System messages are not shown to fit the figure on one page.}
  \label{fig:conversation-breathing-dataset}
\end{figure}

\subsection{Testing the semantic similarity search access rights filter}
To make sure that the semantic similarity search only retrieves vectors accessible to the logged-in user, we constructed several tests. We found that the semantic similarity search only retrieves vectors intended to be accessible to the user. Vectors from records that are private and for which the user has no defined role are not retrieved. Similarly, vectors associated with deleted/inactive records are not found in the set of retrieved records. On the other hand, vectors from records the user may access are retrieved. This includes records that are active (not deleted) and the user has access to, either by a defined role or the record being public.
These findings support that we implemented a semantic similarity search directly in Kadi’s underlying database while respecting Kadi’s fine-grained access-control rules (RQ1).

\subsection{Scalability results} \label{sec:results:scaling}
\subsubsection{Scalability of k-NN search time}
\begin{figure}
    \centering
    \includegraphics[
    alt={A visualization of the k-NN runtime for up to one million vectors for different access fractions. The number of records is fixed at 2.}
    ]{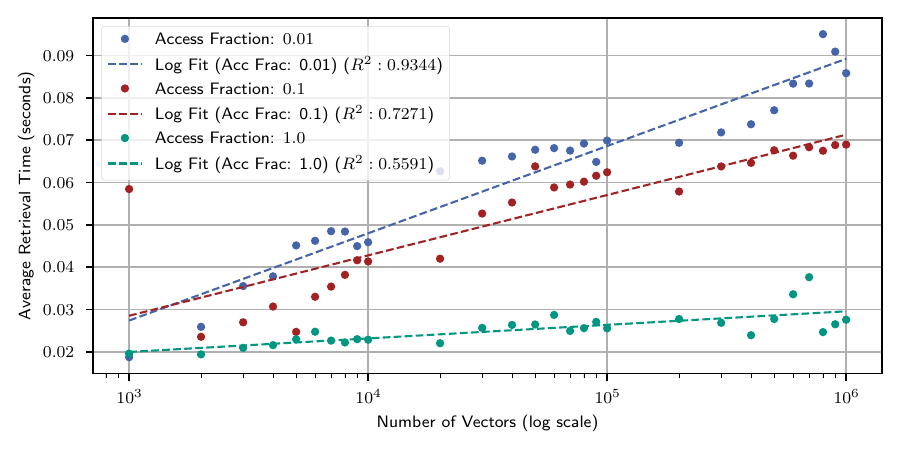}
    \caption{Runtime scaling of the k-NN search for two records. Shown is the average execution time (100 trials per data point) of the k-NN search (k=50) with increasing number of vectors for different access fractions. The number of records is fixed at 2: One record for all accessible vectors and one for all non-accessible.}
    \label{fig:runtime_plot_2_rec}
\end{figure}

To investigate the query time scaling of the k-NN search (RQ2), we conducted two experiments.
\Cref{fig:runtime_plot_2_rec} shows the k-NN runtime results for a fixed number of records as described in \Cref{sec:exp_setup:retrieval_setup}.
The fitted logarithmic functions show good alignment with the measured runtimes.
This observed logarithmic trend in k-NN runtime aligns well with the empirical near-logarithmic query-time scaling of the utilized HNSW index \cite{malkovEfficientRobustApproximate2018}.

\begin{figure}
    \centering
    \includegraphics[
    alt={A visualization of the k-NN runtime for up to one million vectors for different access fractions and different numbers of vectors per record.}
    ]{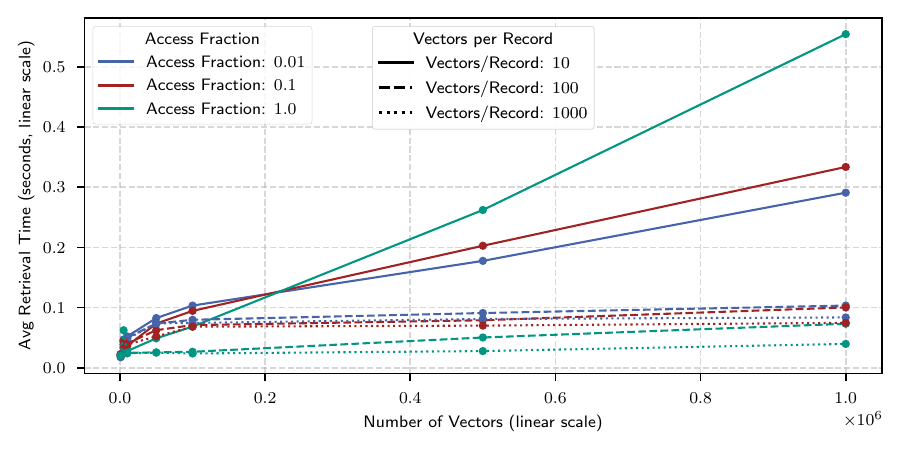}
    \caption{Runtime scaling of the k-NN search with an increasing number of records. Average execution time (500 trials for each data point) of the k-NN search (with k=50) for a growing number of vectors and three access fractions. In this experiment, we define a maximum number of vectors that a record can contain (Vectors/Record). This leads to an approximately linear growth of records with the number of vectors.}
    \label{fig:kadichat_lin_scaling}
\end{figure}

\Cref{fig:kadichat_lin_scaling} shows the results of the second k-NN runtime experiment, for which we increase the number of records approximately linearly with the number of vectors as described in \Cref{sec:exp_setup:retrieval_setup}. Notably, the three configurations with 10 vectors per record scale worse than the others. They show linear scaling of retrieval time, contrary to the approximately logarithmic scaling of an HNSW index \cite{malkovEfficientRobustApproximate2018}. This result shows that the number of records plays a crucial role in the runtime of the k-NN search. Because the access filter for the vectors relies on a query that retrieves all accessible records, this component scales linearly and, in certain configurations, dominates k-NN retrieval time.
This scaling behavior might cause issues with large production Kadi4Mat instances, especially when many records need to be filtered. Nevertheless, retrieval time for most configurations at one million vectors is around \SI{0.1}{\second} or less, which we consider acceptable.

\subsubsection{Memory scalability of the vector database}
\begin{figure}
    \centering
    \includegraphics[
    alt={A visualization of the memory consumption for four different Kadi instances.}
    ]{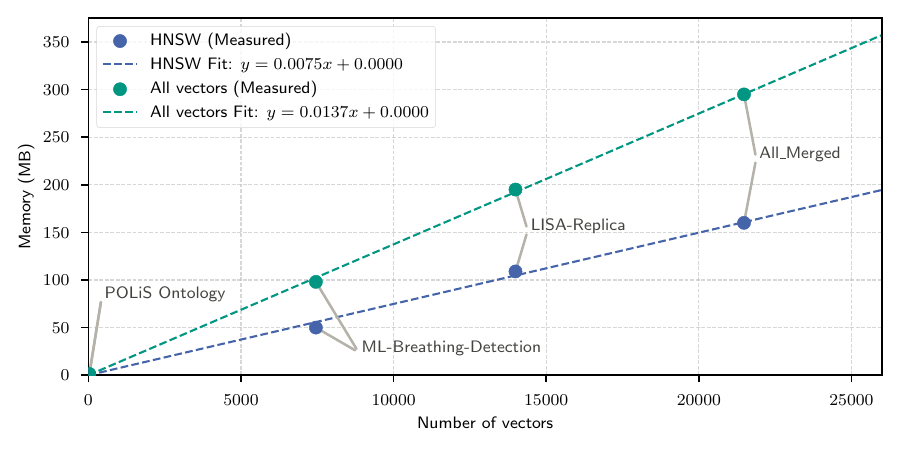}
    \caption{Memory growth of the HNSW index and the total vectors table. The vector table includes the HNSW index and TOAST data. Four Kadi4Mat instances of different sizes were measured.}
    \label{fig:memory_scaling}
\end{figure}

To predict the memory scaling of the vector table in Kadi (RQ2), we constructed four databases and measured their memory consumption as described in \Cref{sec:methods:scalability:memory}.
\Cref{fig:memory_scaling} shows the memory consumption of the HNSW index and the whole vector table for the four databases.
As expected, both the HNSW index and the total vector table memory consumption grow linearly in our experiment. Using the fitted linear functions, we conclude that HNSW index memory consumption roughly scales to \SI{7.5}{\kilo\byte} per vector, with negligible constant overhead. This linear trend results in roughly \SI{7.5}{\giga\byte} for one million vectors, which should fit in the RAM of most Kadi instances. For the vector table's total memory, we found it grows by about \SI{13.7}{\kilo\byte} per vector. For one million vectors, this would be \SI{13.7}{\giga\byte} of disk or SSD space. We conclude that most production instances of Kadi4Mat likely meet the memory needs of the semantic similarity search. 

\subsection{Quantitative answer results} \label{sec:results:quantitative}
To evaluate and compare KadiAssistant and the three baseline models quantitatively, we ran each model five times on our benchmark dataset. As recall, fuzzy precision, strict precision, and source faithfulness produce discrete values (because the sets they are based on are rather small), we visualize their means across all five runs using bar charts with whiskers indicating the minimum and maximum of the five run means. We visualize the metrics answer relevancy and answer correctness using violin plots, with the mean shown as a black bar; again, whiskers show the minimum and maximum means across the five runs. 

\begin{figure}
    \centering
    \includegraphics[
    alt={A visualization of the three source metrics \textit{recall}, \textit{fuzzy precision}, and \textit{strict precision} for KadiAssistant and the three baseline models for each dataset.}
    ]{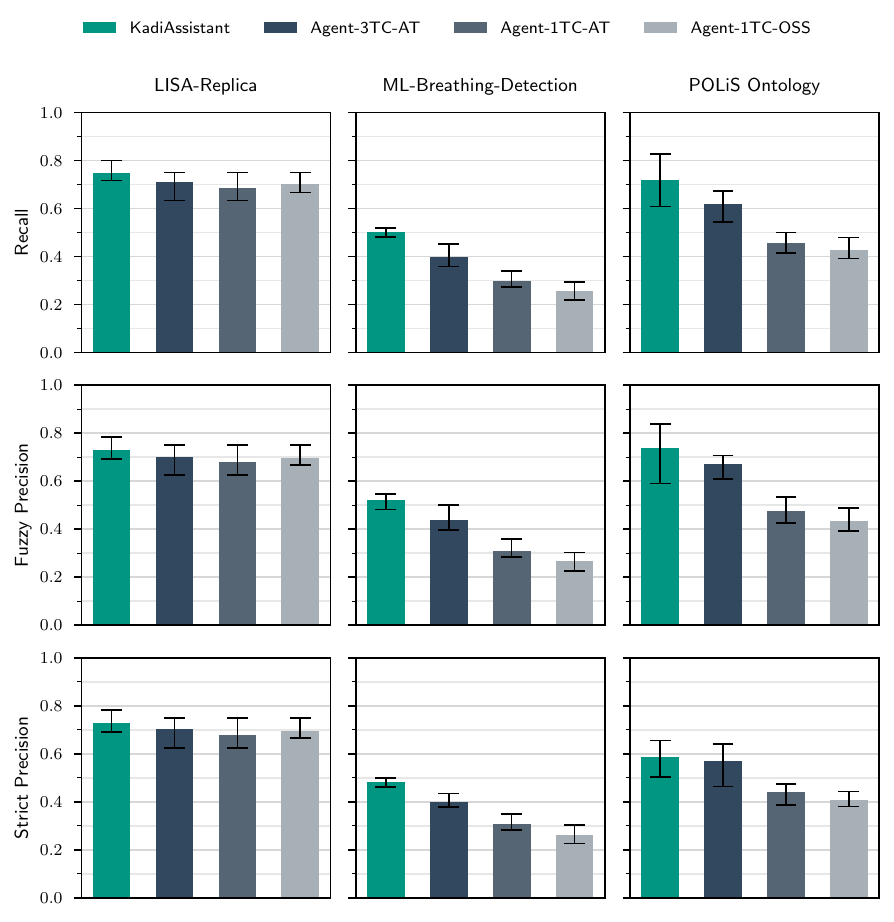}
    \caption{The three source metrics \textit{recall}, \textit{fuzzy precision}, and \textit{strict precision} for KadiAssistant and the three baseline models for each dataset. Whiskers indicate the minimum and maximum mean values across the five executed runs.}
    \label{fig:source_metrics_per_ds}
\end{figure}

\Cref{fig:source_metrics_per_ds} visualizes the three metrics: recall, fuzzy precision, and strict precision for KadiAssistant and the three baseline models separately for each dataset.
For the two ontology-based datasets \textit{ML-Breathing-Detection} and \textit{POLiS Ontology}, we see higher values for KadiAssistant compared to the three baseline models across all three source metrics. For the LISA-Replica dataset, all four models achieve similar scores.

\begin{figure}
    \centering
    \includegraphics[
    alt={A visualization of the three RAGAS metrics \textit{source faithfulness}, \textit{answer relevancy}, and \textit{answer correctness} for KadiAssistant and the three baseline models for each dataset.}
    ]{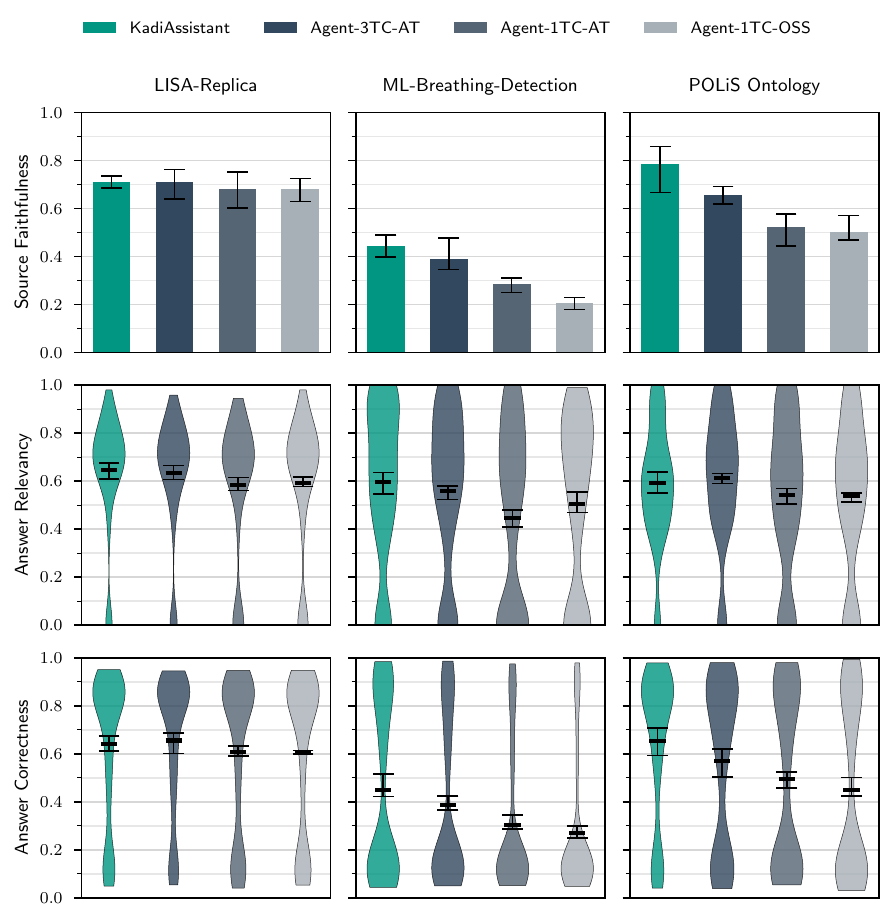}
    \caption{The three RAGAS metrics \textit{source faithfulness}, \textit{answer relevancy}, and \textit{answer correctness} for KadiAssistant and the three baseline models for each dataset. Source faithfulness is shown as bar plots, while answer relevancy and correctness are shown as violin plots. Bars indicate the mean, and whiskers indicate the minimum and maximum mean values across five runs.}
    \label{fig:ragas_metrics_per_ds}
\end{figure}

\Cref{fig:ragas_metrics_per_ds} depicts the RAGAS-based metrics for KadiAssistant and the three baseline models. As with the source metrics, we observe comparable scores for all models on the LISA-Replica dataset. For the two metrics \textit{source faithfulness} and \textit{answer correctness}, we observe better scores for KadiAssistant compared to the single-tool-call baselines on the \textit{ML-Breathing-Detection} and \textit{POLiS Ontology} datasets. For the metric answer relevancy, we did not observe a clear pattern.
Additionally, we found that for all agents, the metrics (except answer relevancy) are considerably lower for ML-Breathing-Detection than for the other two datasets. We assume this is because the ML-Breathing-Detection dataset contains little natural language, and the recorded metadata is very similar across many records, making it difficult for the agents to find the correct resources.

These results address RQ3 and indicate that systems like KadiAssistant with multiple information retrieval steps outperform the single-step baselines on our ontology-based datasets.

\begin{figure}
    \centering
    \includegraphics[
    alt={A visualization of the three source metrics \textit{recall}, \textit{fuzzy precision}, and \textit{strict precision} for KadiAssistant and the three baseline models across question complexities (hop number) in the POLiS Ontology dataset.}
    ]{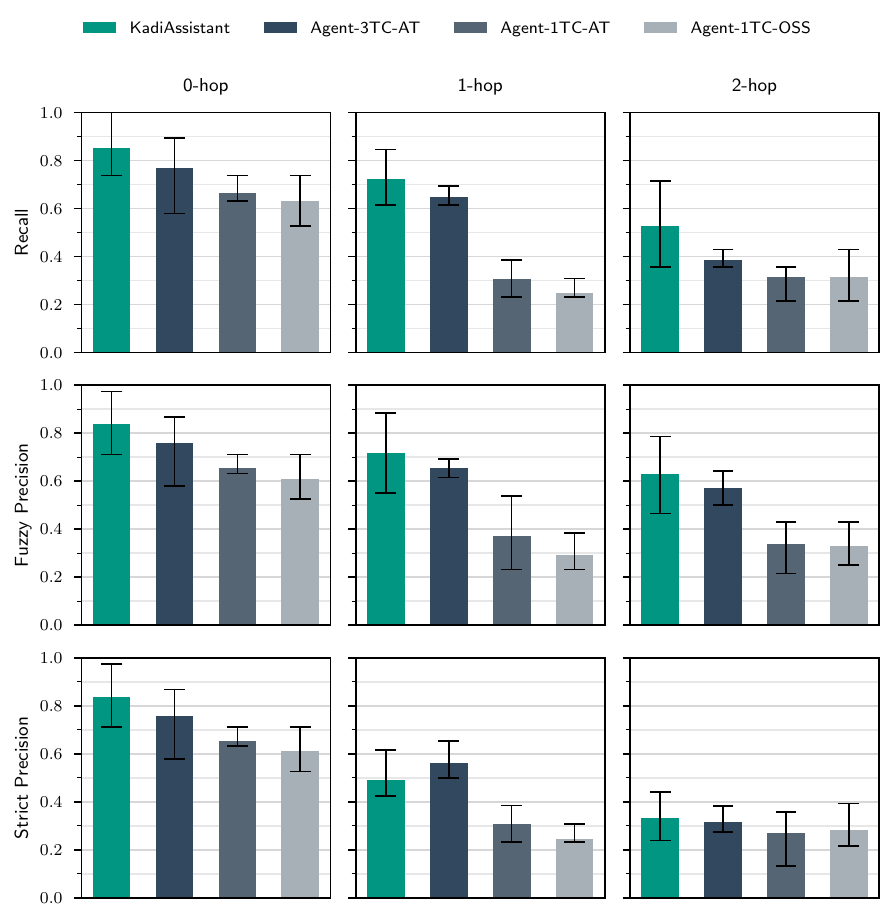}
    \caption{The three source metrics \textit{recall}, \textit{fuzzy precision}, and \textit{strict precision} for KadiAssistant and the three baseline models across question complexities (hop number) in the POLiS Ontology dataset. Whiskers indicate the minimum and maximum mean values across the five executed runs.}
    \label{fig:source_metrics_per_hops_onto4polis}
\end{figure}

\begin{figure}
    \centering
    \includegraphics[
    alt={A visualization of the three RAGAS metrics \textit{source faithfulness}, \textit{answer relevancy}, and \textit{answer correctness}, for KadiAssistant and the three baseline models across question complexities (hop number) in the POLiS Ontology dataset.}
    ]{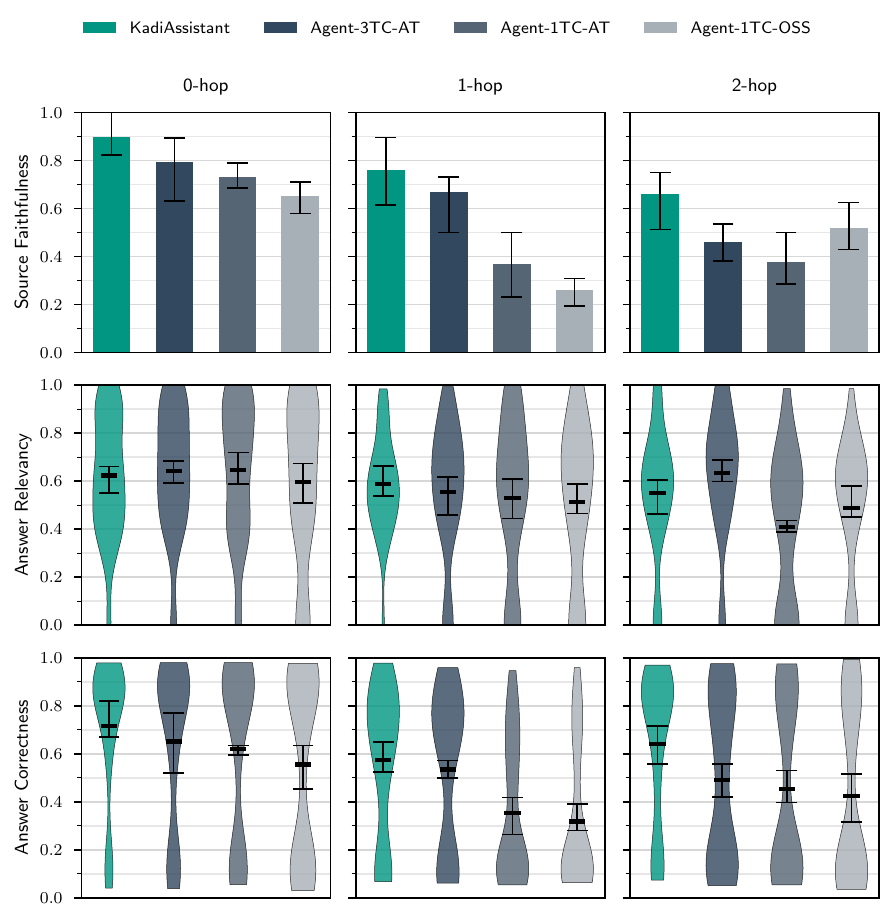}
    \caption{The three RAGAS metrics, \textit{source faithfulness}, \textit{answer relevancy}, and \textit{answer correctness}, for KadiAssistant and the three baseline models across question complexities (hop number) in the POLiS Ontology dataset. Source faithfulness is shown as bar plots, while answer relevancy and correctness are shown as violin plots. Bars indicate the mean, and whiskers indicate the minimum and maximum mean values across five runs.}
    \label{fig:ragas_metrics_per_hops_onto4polis}
\end{figure}

To address part two of RQ3, we analyzed the performance of the assistants for different hop counts for each dataset. \Cref{fig:source_metrics_per_hops_onto4polis} and \Cref{fig:ragas_metrics_per_hops_onto4polis} depict the source and RAGAS metrics on the POLiS Ontology dataset for each hop count. KadiAssistant clearly outperforms the baseline models for the recall, fuzzy precision, source faithfulness, and answer correctness metrics across all hop counts. Interestingly, from 0-hop to 1-hop, the recall and fuzzy precision scores decrease significantly more for the single-tool-call baseline models than for KadiAssistant and the Agent-3TC-AT baseline model. From 1-hop to 2-hop, however, the scores for the single-tool-call baselines stay roughly the same, while the scores for the multi-step assistants, KadiAssistant and Agent-3TC-AT, decrease significantly. Even so, KadiAssistant and the Agent-3TC-AT baseline still outperform the single-tool-call baselines at 2-hop questions. Why the Agent-1TC-AT and Agent-1TC-OSS baseline scores plateau rather than decline remains unclear. The supplementary material (Section 3) details the metrics on the other two datasets.

In addition to the metrics, we tracked further statistics during evaluation (\Cref{tab:eval-stats}). In rare cases (<1\%), the multi-step-retrieval systems threw an error and did not return an answer. We also found that some assistant-generated sources were malformed because they either did not follow the defined JSON format or contained resource identifiers that could not be found on the Kadi instance. Consequently, malformed sources cannot match any source defined in gt-full-sources or gt-answer-sources and are therefore handled as false positives.
We also noticed that several RAGAS metric values are missing (caused by errors), with 3-7\% missing values for source faithfulness, up to 4\% for answer correctness, and under one percent for answer relevancy. We handled the missing RAGAS-based metrics for our evaluation plot by discarding them. These missing values may influence the shown results.
We assume that a considerable fraction of the missing source faithfulness values result from the context window being exceeded when large records are included in the generated sources.

\begin{table}[htbp]
\centering
\caption{Generation and evaluation error statistics across configurations, for a total of 795 queries per assistant (159 questions, five runs).} 
\label{tab:eval-stats}
\begin{tabular}{lcccc}
\toprule
 \textbf{Metric}& \textbf{KadiAssistant} & \textbf{Agent-3TC-AT} & \textbf{Agent-1TC-AT} & \textbf{Agent-1TC-OSS} \\
\midrule
Errors during answer process        & 3 (<1\%)  & 7 (<1\%)  & 0 (0\%)  & 0 (0\%)  \\
Answers with malformed sources         & 23 (3\%) & 28 (4\%) & 13 (2\%) & 16 (2\%) \\
Total malformed sources             & 23 & 29 & 13 & 16 \\
Missing source faithfulness values         & 55 (7\%)& 40 (5\%) & 25 (3\%) & 36 (5\%) \\
Missing answer correctness values   & 13 (2\%)& 7 (<1\%)  & 13 (2\%) & 29 (4\%) \\
Missing answer relevancy values     & 4  (<1\%)& 7 (<1\%)  & 0 (0\%)  & 0 (0\%)  \\
\bottomrule
\end{tabular}
\end{table}

\subsection{Limitations and practical challenges of KadiAssistant} \label{sec:limitations_and_challenges}
Despite the positive results and demonstration of KadiAssistant, the system comes with limitations and practical challenges. Sometimes KadiAssistant does not find the intended information, or it can, like all LLM-based AI agents, make mistakes or hallucinate \cite{huangSurveyHallucinationLarge2025}.

\paragraph{Hallucination and unintended behavior.}
Quantitatively, benchmark evaluations show that KadiAssistant may not always find the desired information. On the ML-Breathing-Detection dataset, for example, recall is only about 0.5, with low precision and a low source faithfulness score below 0.5. KadiAssistant may also return malformed or incorrect resources, as shown in \Cref{tab:eval-stats}. Whether these results are partly attributable to hallucinations requires further investigation.

Qualitatively, we also found failure cases:
One failure example is tied to the user question ``Can you locate the paper `Kadi4Mat: A research data infrastructure for materials science' for me?'' (depicted in \Cref{fig:conversation-lisa-kadi}). Occasionally, KadiAssistant referenced the wrong paper, even though the correct paper was cited in its sources. 

We also found that the semantic similarity search sometimes failed for tasks on the ML-Breathing-Detection dataset. In this dataset, the metadata for many records are very similar and sometimes differ only in a few values. Similarly, we identified an issue in the AI assistant's decision chain. By tracing the connection between one machine learning model and a dataset record, KadiAssistant inferred that all models had been trained using this dataset. In this case, the conclusion is correct, but if there were other datasets not tied to the investigated machine learning model, the AI assistant's response would be incorrect. 

Another challenge we faced is that KadiAssistant sometimes fails to assess its tools' capabilities correctly. In these cases, the assistant seems to assume that the semantic similarity search information is complete, even though some information is missing. This happens when the user submits a query that requires analyzing many records, e.g., when asking for the best-performing model (each model is stored in a record). As the semantic similarity search only retrieves a fixed number of chunks (eight in our configuration), not all necessary records may be retrieved. However, KadiAssistant does not recognize this limitation and derives the wrong solution from the chunks. This behavior persists even when adding the limitations of the tools to the corresponding descriptions.

To mitigate the harmful consequences of unintended behavior and hallucination, KadiAssistant supplies the user with sources that the user can use to verify the answer. However, this can be difficult in certain cases; for example, when KadiAssistant incorrectly assumes the completeness of the information it found. In such cases, the AI assistant's answer may be backed by sources but still be incorrect, because it did not find and evaluate all necessary sources.

\paragraph{Prompt injection.}
Another possibility that may lead to unintended behavior is prompt injection. If data stored on Kadi4Mat intentionally or unintentionally contains text that can be understood as instructions for KadiAssistant, it may cause unintended behavior. This corresponds to the problem of indirect prompt injection, in which instructions embedded in the retrieved content are processed by the LLM and may cause it to deviate from the intended task \cite{greshakeNotWhatYouve2023}.
In our architecture and experiments, we did not address this issue directly, and such cases may occur in practice.
However, KadiAssistant's current capabilities, limited to information retrieval, reduce the damage prompt injection can cause. Potential consequences for KadiAssistant include incorrect answers or sources, unintended tool calls, or failure to provide an answer. However, with its current toolset, KadiAssistant cannot alter or delete data stored on Kadi or send it to unauthorized parties.

\paragraph{Scaling.}
A practical challenge may arise as KadiAssistant scales to larger Kadi instances, as k-NN query time may scale linearly for certain configurations. In our experiments, this happened in configurations where vectors per record are set to 10. This scaling might require optimizing the access rights filter for the vectors in the future.

\section{Concluding Remarks} \label{sec:conclusion}
In this work, we presented KadiAssistant, our AI agent for conversational information retrieval in Kadi4Mat. We developed an agentic architecture that combines semantic similarity search with record-metadata and link retrieval in a research data infrastructure with fine-grained access control. We provide qualitative demonstrations and quantitative results addressing the posed research questions.

For \emph{RQ1}, we showed that semantic similarity search can be implemented directly in Kadi4Mat's database while respecting the repository's access-control rules. We supported this claim with several tests showing that the similarity search only returns vectors the logged-in user may access.

For \emph{RQ2}, we found that k-NN query time depends not only on the number of vectors but also on the number of records. We also found that the memory scales roughly linearly with the number of vectors, as expected. We concluded that for most configurations, query time and memory consumption should scale to larger Kadi instances. For certain configurations, however, the query times scale steeply linearly, which may cause problems in the future.

For \emph{RQ3}, our results indicate that KadiAssistant, a multi-step retrieval AI agent, improves answer and source scores compared to single-step-retrieval agents on our ontology datasets across all question complexities. On the LISA-Replica dataset, however, all models performed similarly.

Despite the positive benchmark results and examples, we found some limitations typical of LLM applications. In some cases, the LLM hallucinates or makes errors, especially when information retrieved by the semantic similarity search is misleading, incomplete, or irrelevant. We also identified potential challenges for query-time scaling of the vector database for certain configurations.

Next, we plan to address these issues in KadiAssistant and deploy it to production Kadi4Mat instances. To do so, we will test the system performance with larger databases and higher levels of user activity. Moreover, we have to ensure that the system, especially the Kadi core functionality, remains operational under stress and, if necessary, optimize the performance of some components.
After a first rollout and potential performance improvements, we plan to add more features. First, we want to focus on the information gathering capabilities of KadiAssistant, which include:
\begin{itemize}
    \item An expansion and restructuring of the vector database to support data from collections, templates, users, and more.
    \item Adding filters to the semantic similarity search, so that the AI agent can focus its search on the most relevant Kadi4Mat resources (records, collections, files).
    \item A new data querying tool enabling exact metadata search.
    \item Unification of the existing Kadi search functionality and the new semantic similarity search. 
    \item Data analysis and file summarization tools to enable deeper analysis of tables and better understanding of text files. 
\end{itemize}
After improving KadiAssistant's information gathering capabilities, we plan to make the system accessible across different work environments, e.g., labs, where hands-free use is beneficial. 

Furthermore, we plan to expand KadiAssistant's objective beyond information retrieval to support good research data management. This assistant will help researchers upload and structure their data uniformly.

\newpage

\printbibliography

\section*{Acknowledgments}
This work contributes to the research performed at CELEST (Center for Electrochemical Energy Storage Ulm-Karlsruhe) and was funded by the Deutsche Forschungsgemeinschaft (DFG, German Research Foundation) under Germany’s Excellence Strategy – EXC 2154 – Project ID 390874152 (POLiS Cluster of Excellence).

Contributions are provided through the ``Materials Science and Engineering (MSE)'' programme No. 43.31.01, supported by the Helmholtz Association, which is gratefully acknowledged. 

Supported by the Deutsche Forschungsgemeinschaft (DFG, German Research
Foundation) through CRC 1574 — Project 471687386 and RU 5966 – 556363981

The authors Selzer and Islam would like to thank the Federal Government and
the Heads of Government of the Länder, as well as the Joint Science Conference
(GWK), for their funding and support within the framework of the NFDI4ING
consortium. Funded by the German Research Foundation (DFG) - project number
442146713.

This publication was supported by the Helmholtz Metadata Collaboration (HMC),
an incubator-platform of the Helmholtz Association within the framework of the
Information and Data Science strategic initiative, within the MetaSurf
project.

The authors gratefully acknowledge the data contribution of Manuel Dillenz and Sebastian Baumgart to the POLiS Ontology dataset. 

\section*{Use of AI Tools}
For this work, we used AI tools to improve the paper and code.
The following is a list of AI Tools and their applications in this work.
Note that we cannot guarantee the completeness of the list.
\begin{itemize}
    \item Code:
    \begin{itemize}
        \item Frontend (chatbox-form): Large portions were created or refactored using Copilot (several different models); the code was checked and adapted by the authors
        \item Docstrings were partially generated using the autocomplete feature of Copilot and checked and improved afterward by the authors 
        \item Parts of the code were generated using the autocomplete feature of Copilot and checked and improved afterward by the authors 
        \item Parts of the prompting for the benchmark generation were generated by Claude and reviewed by the authors
        \item The helper script for de-duplication (\Cref{sec:methods:benchmark_construction}) was generated using Claude
    \end{itemize}
    \item Paper:
    \begin{itemize}
        \item ChatGPT (various models) and Grammarly were used to improve the readability and flow of the paper. Content comes from the authors; AI-enhanced texts were checked and adapted by them.
        \item Claude Code was used to generate large portions of the code necessary for the illustration of chat messages in \Cref{sec:demo}.
    \end{itemize}
\end{itemize}

\section*{Availability of Data and Code}
Code for the KadiAssistant components: 
\begin{itemize}
    \item Kadi with semantic similarity search:\\https://gitlab.com/intelligent-analysis/kadiai/kadichat2.0/kadi-vectors
    \item The AI agent component:\\https://gitlab.com/intelligent-analysis/kadiai/kadichat2.0/kadichat2.0
    \item The agentstoolbox, containing the tools for the AI agent:\\ https://gitlab.com/intelligent-analysis/kadiai/kadichat2.0/agentstoolbox
\end{itemize}
Code For the Experiments:
\begin{itemize}
    \item The code for the scaling experiments:\\https://gitlab.com/intelligent-analysis/kadiai/kadichat2.0/experiments4vectordb
    \item The code for the benchmark and quantitative analysis:\\
    https://gitlab.com/intelligent-analysis/kadiai/kadichat2.0/kadiassistant-benchmark
\end{itemize}
All code can also be found on Zenodo: \cite{cierpka_2026_22766721}

The data for the POLiS Ontology dataset and the benchmark evaluation can be found on Zenodo \cite{cierpka_2026_22814081}.
We did not publish the LISA-Replica and ML-Breathing-Detection datasets due to copyright and data sensitivity concerns.
\end{document}